\documentstyle[prc,aps]{revtex}
\def\vec#1{\bbox{#1}}
\begin{document}
\title{
On Nonlinear Evolution of Axisymmetric Nuclear Surface
}
\author{
V.G. Kartavenko$~^{1,2}$,
K.A. Gridnev$~^3$ and W.Greiner$~^2$
}
\address{
$^1$~Bogoliubov Laboratory of Theoretical Physics,
Joint Institute for Nuclear Research,\\
Dubna, Moscow District, 141980, Russia\\
$^2$~Institut f\"ur Theoretische Physik der J. W. Goethe Universit\"at\\
D-60054 Frankfurt am Main, Germany\\
$^3$~Department of Nuclear Physics,Institute of Physics,\\
St.Petersburg State University, 198904, Russia
}
\date{November 2000}
\maketitle
\begin{abstract}
We consider an uniformly charged incompressible nuclear
fluid bounded by a closed surface.
It is shown that an evolution of an axisymmetric surface
$\Gamma(\bbox{r},t)\equiv \sigma - \Sigma(z,t) = 0,\quad
\bbox{r}=(\sigma,\phi,z)$ can be approximately reduced to
a motion of a curve in the $(\sigma,z)$-plane.
A nonlinear integro-diffrerential equation for the contour
$\Sigma(z,t)$ is derived. It is pointed on
a direct correspondence between
$\Sigma(z,t)$ and a local curvature, that gives possibility to
use methods of differential geometry to
analyze an evolution of an axisymmetric nuclear surface.
\end{abstract}
\section{Motivation}
Nonlinear dynamics of a nuclear surface is an object
of special interest due to the following reasons.
First, nuclear density falls considerably in the surface region,
where the density fluctuations and clustering may be important.
Second, different types of instability may be developed
in the surface region and lead to fragmentation processes
at low (fission, nucleon transfer) and high (multifragmentation,
break-up etc.) energies.
Finally,
the liquid drop concept~\cite{RayleighLD}  for over
a century are intensively used in macro-~\cite{chandra} 
and micro-physics~~\cite{EisGr}.
Nonlinear dynamics of shapes in any complicated systems
enevitably leads to mathematical problem of describing
global geometric quantities such as surface and enclosed
volume in different dimensions (polymers, cell membranes,
3D droplets).

An application of a soliton concept to
nonlinear nuclear hydrodynamics has given yet new
possibilities in this field (See e.g. review~\cite{kar93pn}
and ~\cite{ApfelPRL97,ludraa98,kgg98,ksg99} for the recent refs.).
However any extension
of nonlinear dynamics from 1+1 to 2+1 and 3+1 dimensions
meets principal difficulties. The crucial point  is to
reduce the dimension of a problem. In paper~\cite{kartJP93}
we considered the simplest two-dimensional nonlinear liquid objects.
It was shown that 2D pure vortical motion of inviscid nuclear
liquid can be reduced to 1D evolution of the contour bounding this
drop.
In the next paper~\cite{KGMG96} the extension
to semi-3D geometry was done. The equations of motion describing
localized vortex on a spherical nuclear surface -
a bounded region of constant vorticity, surrounded by irrotational
flux - were reduced to 1D nonlinear evolution of the boundary.

In this short report
we consider an uniformly charged incompressible nuclear 3D
fluid bounded by a closed surface.
It is shown that evolution of an axisymmetric surface
$\Gamma(\vec{r},t)\equiv \sigma - \Sigma(z,t) = 0,\quad
\vec{r}=(\sigma,\phi,z)$ can be approximately reduced to
a motion of a curve in the $(\sigma,z)$-plane.

\section{Framework}
The evolution of one-body
Wigner phase-space distribution function is analyzed instead of a full
many-body wave function. Integrating the kinetic equation
\begin{equation}
\frac{\partial f}{\partial t}+\frac{\bbox{p}}{m}\cdot
\frac{\partial f}{\partial{\bbox r}}-\frac{\partial V}{\partial
{\bbox r}}\cdot
\frac{\partial f}{\partial\bbox{p}}=I_{rel},\qquad
H_{W} = \frac{p^{2}}{2m} + V(\bbox{r})
\label{vlasov}
\end{equation}
over the momentum space with different
polynomial weighting functions of the ${\bbox p}$-variable
one comes to an infinite chain of equations
for local collective observables including the density, collective
velocity,
pressure and an infinite set of tensorial functions of the time and space
coordinates, which are defined as moments of the distribution function in
the momentum space:\\
- the particle $n({\bbox r},t)\equiv g\int d\bbox{p}\,f({\bbox r},\bbox{p},t),$
and the mass $\rho({\bbox r},t)=m\,n({\bbox r},t)$ densities,
where we consider a proton and neutron as different state of identical
particle (the spin-isospin degeneration $g = 4$).\\
- the collective current and velocity of nuclear matter
$\rho ({\bbox r},t){\bbox  u}({\bbox r},t)=
g\int d\bbox{p}\,\bbox{p}f({\bbox r},\bbox{p},t)$\\
- the pressure tensor
$P_{ij}({\bbox r},t)=({g}/{m}) \int d\bbox{p}\, q_{i}
q_{j}f({\bbox r},\bbox{p},t),\quad
{\bbox q}=\bbox{p}-m{\bbox  u}$\\
- the energy and momentum transfer tensors of a different order
\begin{eqnarray}
P_{ijk}({\bbox r},t)=\frac{g}{m^{2}}\int d\bbox{p}\, q_{i}
q_{j} q_{k} f({\bbox r},\bbox{p},t),\qquad
P_{\underbrace{ij..k}_{n}}({\bbox r},t)=
\frac{g}{m^{n-1}}\int d\bbox{p}\,
\underbrace{q_{i}q_{j}..q_{k}}_{n} f({\bbox r},\bbox{p},t),
\end{eqnarray}
- and the  integrals related to relaxation terms
\begin{equation}
\int d\bbox{p}\,I_{rel}=0,\qquad
\int d\bbox{p}\, \bbox{p} I_{rel}=0,\qquad
R_{ij} \equiv \frac{g}{m}
\int d\bbox{p}\,q_{i}q_{j}I_{rel},\quad\dots
\end{equation}
Truncating  this chain
one arrives at the "fluid
dynamical" level of description of nuclear processes.
\begin{equation}
\frac{\partial \rho}{\partial t} + \sum\limits_{k}
\frac{\partial}{\partial x_{k}} (u_{k}\rho) = 0,
\label{hyd1}
\end{equation}
\begin{eqnarray}\nonumber
\rho \frac{D u_{i}}{D t}  &+& \sum\limits_{k}
\frac{\partial P_{ik}}{\partial x_{k}}
+ \frac{\rho}{m} \frac{\partial V}{\partial x_{i}}
+ \rho (\Omega_{i} \sum\limits_{k}\Omega_{k}x_{k} - \Omega^{2} x_{i} )\\
&+& \rho \sum\limits_{s,j} \varepsilon_{isj} (2\Omega_{s} u_{j} +
\frac{d\Omega_{s}}{d t}x_{j}) = 0,
\label{hyd2}
\end{eqnarray}
\begin{eqnarray}\nonumber
\frac{D P_{ij}}{D t} &+& \sum_{k}\left( P_{ik}
\frac{\partial u_{j}}{\partial x_{k}} + P_{jk}
\frac{\partial u_{i}}{\partial x_{k}} +
P_{ij}\frac{\partial u_{k}}{\partial x_{k}}\right) \\
&+& 2\sum\limits_{s,m} \Omega_{m} (\varepsilon_{jms} P_{is} +
\varepsilon_{ims}P_{js}) +
\sum\limits_{k}\frac{\partial}{\partial x_{k}}P_{ijk} = R_{ij}
\label{hyd3}
\end{eqnarray}
where the usual notation
$\frac{D}{D t}\equiv \frac{\partial }{\partial t}
+ \sum\limits_{k} u_{k} \frac{\partial}{\partial x_{k}}$
is introduced for the operator giving the material derivative,
or rate of change at a point moving with the fluid locally.
The hydrodynamical set of Eqs.~(\ref{hyd1}-\ref{hyd3})
describes an evolution in a rotating nuclear system.
The linear transformation
$x_{i} = \sum\limits_{j=1}^{3} T_{ij} X_{j}$
relates the coordinates of a point, ($X_{1}, X_{2}, X_{3}$), in 
an inertial frame, and,($x_{1}, x_{2}, x_{3}$ in a moving frame
of reference with a common origin.
The orientation of a moving frame,
with respect to an inertial frame, will be assumed to be time dependent,
$T_{ij}(t)$, representing an orthogonal transformation, and
the vector
\[
\Omega_{i} = \frac{1}{2}\sum_{j,k,m}\varepsilon_{ijk}
\left({dT}/{d t}\right)_{jm} T^{+}_{mk}
\]
represents a general time-dependent rotation.

Let us restrict ourselves with simplest possible motion of nuclear fluid.
Consider an uniformly charged (with a total charge $Ze$)
incompressible nuclear "fluid" confined to a volume $V$
bounded by a closed surface $S$, obeying to the equation
$\Gamma(\bbox{r},t)\equiv \sigma - \Sigma(z,t) = 0$, where
$(\sigma,\phi,z)$ are cylindrical coordinates of a point.
The geometry places the origin at the center-of-mass
($\int d\bbox{r}\,\rho ({\bbox r},t){\bbox  u}({\bbox r},t)=0$).
The mean-field potential and related tensors can be decomposed into
nuclear and Coulomb terms $V({\bbox x}, t) = V_{nucl}({\bbox x}, t) +
V_{coul}({\bbox x}, t)$.
A nuclear potential can be derived as the first variation of
the short-range interaction functional
(as usual effective density dependent Skyrme forces) on density
$V_{nucl}({\bbox x}, t) \equiv \delta{\cal E} [ n ]/\delta n$,
This gives a nuclear potential as the function of
the density $n(\bbox{r},t)=n_{0}\eta(\bbox{r},t)$.
The incompressibility of the nuclear matter implies 
$n_{0} = (2p_F^3)/(3\pi\hbar^3)$
where $n_{0},\;p_{F}$
being respectively,
the density of nuclear matter  and the Fermi momentum.
Or, keeping in mind that an effective mean-field potential approximately
constant inside the nuclei and has a sharp coordinate dependence in the
surface region, to replace it
by the surface term\cite{rosenkilde}
$\gamma_{s} div\hat{\bbox{n}}$,
where
$\hat{\bbox{n}} = \nabla\Gamma/|\nabla\Gamma|$
is the unit outward normal on $S$, and
the surface tension $\gamma_{s}$ is related to mass formula 
coefficient via
$b_{s} = 4\pi r_{0}^{2}\gamma_{s} \sim 22 MeV$.
In the both cases the contour $\Sigma(z,t)$ defines completely
a nuclear potential on a nuclear surface.
\begin{eqnarray}\nonumber
 V_{coul}({\bbox x}, t)= \left(\frac{Z}{A} e\right)^{2} n_{0}
\int d\bbox{x'} \, \frac{\eta(\bbox{x}', t)}{|\bbox{x}-\bbox{x} '|} =\\
\nonumber
 \left(\frac{Z}{A} e\right)^{2} n_{0}
\int_{z_{min}}^{z_{max}}\;dz' \int_{0}^{\Sigma(z',t)} d\sigma' \sigma'
I(\sigma, \sigma', z'),\\
I(\sigma, \sigma', z') = \int_{0}^{2\pi} d\phi
\frac{1}{\sqrt{\sigma^{2}+\sigma'^{2}-2\sigma\sigma'\cos\phi}},
\label{coul}
\end{eqnarray}
%
Concerning pressure tensors,
we will not solve the kinetic equation (\ref{vlasov}),
but  will use that,
this simplest picture can be reproduced by assuming
the following factorised form of
the distribution functions $f(\bbox{r},\bbox{p},t)$
\begin{eqnarray}
f^{0}(\bbox{r},\bbox{p})=(2\pi\hbar)^{-3}\eta^{0}(\bbox{r})
\;\phi^{0}(\bbox{p})\qquad
f(\bbox{r},\bbox{p},t)= (2\pi\hbar)^{-3}\eta (\bbox{r},t)\;
\phi(\bbox{r},\bbox{p},t),
\end{eqnarray}
where $\eta(\bbox{r},t)\equiv\theta(\Gamma(\bbox{r},t))$
($\eta^{0}(\bbox{r})\equiv\theta(\Gamma^{0}(\bbox{r}))$)
is the step-function which is
equal to one inside the surface $\Gamma(\bbox{r},t)$
($\Gamma^{0}(\bbox{r})$)and zero
outside it. The distribution function $f^{0}(\bbox{r},\bbox{p})$
describes the equilibrium spherical symmetrical state
$(\Gamma^{0}(\bbox{r}))\equiv\sigma-\sqrt{R_{0}^{2}-z^{2}})$,
where $R_{0}$ is the radius of a sphere of an equivalent volume
($4R_{0}^{3}=3\int dz \Sigma(z,t)^2$).
The distribution
function $f(\bbox{r},\bbox{p},t)$ corresponds to the dynamical picture.
Both distribution function describe a homogenous distribution
of a nuclear matter within the volume bounded by a sharp surface.
The momentum-dependent part of the distribution functions could be
parametrized as follows :
\begin{eqnarray}
\phi^{0}(\bbox{p})=\theta(p^{2}_{F}-p^{2}),\qquad
\phi (\bbox{r},\bbox{p}, t)=\theta(p^{2}_{F}-
\sum_{i,j}(\delta_{ij}+\alpha_{ij} (t)) q_{i}q_{j}).
\label{P_NoS}
\end{eqnarray}
The forms chosen for the momentum-dependent
part of $f(\bbox{r},\bbox{p},t),\;(f^{0}(\bbox{r},\bbox{p},t))$,
insures that it yields the current
density with the collective velocity $\bbox{u}(\bbox{r},t)
(\bbox{u}^{0}=0)$. Of course, symmetry of the problem is defined
by the symmetry of the all effective forces and an initial
state.
In a fluid at rest, only normal stresses are exerted,
the normal stress is independent of the direction of the normal
to the surface element across which it acts, and
the equilibrium pressure tensor
has the spherically symmetrical form via symmetry of the equilibrium
distribution function
$P^{0}_{ij}({\bbox r}, t) = ({2}/{5}) \epsilon_{F}
 \eta^{0}({\bbox{r}}) n_{0} \delta_{ij}$.
There no reason to expect these results to be valid for
a fluid in motion, the tangential stresses are non-zero, in general, and
the normal component of the stress acting across a surface element
depends on the direction of the normal to the element.
The simple notion of a pressure acting equally in all
directions is lost in most cases of a fluid in motion.
A deformations in the cartesian space define
deformation in the momentum space and vice versa.
We consider the only motion with $\bbox{\Omega}=0$ and will
assume
the diagonal tensor $\alpha_{ij}$.
The dynamical pressure tensors are
$P_{ij}({\bbox r}, t) =
(2/5)(1+\alpha_{ii})^{-1}
\epsilon_{F}n_0 \eta({\bbox{r},t})\delta_{ij}$
These equations  show that the introduction of the functions $\alpha_{ii}(t)$
into the expression for $\phi (\bbox{r},\bbox{p},t)$
accounts for the deformation of the Fermi-surface.

The main hydrodynamical set of equations
can be recast in the following form in the cylindrical coordinates
\begin{eqnarray}\nonumber
\frac{\partial u_{\sigma}}{\partial t} &+&
u_{\sigma}\frac{\partial u_{\sigma}}{\partial\sigma} +
u_{z}\frac{\partial u_{\sigma}}{\partial z} = F_{\sigma},\\
\nonumber
\frac{\partial u_{z}}{\partial t} &+&
u_{\sigma}\frac{\partial u_{z}}{\partial\sigma} +
u_{z}\frac{\partial u_{z}}{\partial z} = F_{z},\\
\frac{\partial u_{\sigma}}{\partial\sigma} &+&
\frac{\partial u_{z}}{\partial z} +
\frac{u_{\sigma}}{\sigma} = 0,
\label{hyd-cyl}
\end{eqnarray}
where
$u_{\sigma},\;u_{z}$
($F_{\sigma},\;F_{z}$) are the projections of
the velocity (forces) along
the axis $\sigma$ and  $z$ respectively.

The boundary conditions are
\begin{equation}
\hat{\bbox{n}} \Xi \hat{\bbox{n}} = - \gamma_{s}
\bigl( \frac{1}{C_{1}}+\frac{1}{C_{2}} \bigr),\qquad
\hat{\bbox{n}} \Xi \hat{\bbox{t}} = 0.
\label{bc_nn}
\end{equation}
Here $\Xi$ is the stress tensor, which can be calculated
using all the formulas above for the pressure tensor and
the mean-field potentials.
$C_{1}$ and $C_{2}$ are the principal radii of curvature.
The direct evaluation gives 
\begin{equation}
\frac{1}{C_{1}}+\frac{1}{C_{2}}
= \frac{1}{\Sigma
(1+\Sigma_z^2)^{1/2}} -
\frac{\Sigma_{zz}}
{(1+\Sigma_z^2)^{3/2}},\qquad\Sigma_z\equiv
\frac{\partial\Sigma}{\partial z},\qquad
\Sigma_{zz}\equiv\frac{\partial^2\Sigma}{\partial z^2}
\label{totcurv}
\end{equation}
\begin{eqnarray}
\hat{\bbox{n}} = \frac{\hat{e}_{\sigma}-
\Sigma_z\hat{e}_{z}}
{\sqrt{1+\Sigma_z^2}},\qquad
\hat{\bbox{t}} = \frac{\Sigma_z\hat{e}_{\sigma}
+\hat{e}_{z}}
{\sqrt{1+\Sigma_z^2}}\qquad
\hat{e}_{\sigma} = \frac{\hat{\bbox{n}}+
\Sigma_z\hat{\bbox{t}}}
{\sqrt{1+\Sigma_z^2}},\qquad
\hat{e}_{z} = \frac{-\Sigma_z
\hat{\bbox{n}}+\hat{\bbox{t}}}
{\sqrt{1+\Sigma_z^2}},
\label{nt_P}
\end{eqnarray}
The final integro-differential equation of motion for a contour
 accumulates the coupled set of equations
(\ref{hyd-cyl}-\ref{totcurv}) 
\begin{equation}
\frac{D\Sigma(z,t)}{Dt} =
\frac{\partial \Sigma(z,t)}{\partial t} +
u_{z}(\Sigma,z,t)\frac{\partial \Sigma(z,t)}{\partial z} =
u_{\sigma}(\Sigma,z,t)
\label{contour}
\end{equation}
The evolution of the surface $\Gamma(\sigma, z, t)$ as
a motion of a 2D space curve in the $(\sigma,z)$-plane is
a particular choice of a general 3D curve dynamics
\begin{equation}
\bbox{u} =
u_{t}\hat{\bbox{t}} +
u_{n}\hat{\bbox{n}} +
u_{b}\hat{\bbox{b}}
\label{fs1}
\end{equation}
where the tangent $\hat{\bbox{t}}$, normal
$\hat{\bbox{n}}$, and binormal
$\hat{\bbox{b}}\equiv\hat{\bbox{t}} \times \hat{\bbox{n}}$
comprise the Frenet-Serret triad\cite{Smirnov57},
obeying the equations of differential geometry
\begin{eqnarray}\nonumber
\hat{\bbox{t}} \equiv \frac{d\bbox{r}}{ds},\qquad
\frac{d\hat{\bbox{t}}}{ds} = \frac{d^{2}\bbox{r}}{ds^{2}} =
\kappa\hat{\bbox{n}},  \qquad
\kappa = | \frac{d^{2}\bbox{r}}{ds^{2}} |\\ 
\frac{d\hat{\bbox{b}}}{ds} = -\tau\hat{\bbox{n}},\qquad
\tau = \kappa^{-2}\Bigl(
\frac{d\bbox{r}}{ds} \times
\frac{d^{2}\bbox{r}}{ds^{2}}\Bigr)\cdot
\frac{d^{3}\bbox{r}}{ds^{3}}\qquad
\frac{d\bbox{n}}{ds} = -
\kappa\hat{\bbox{t}} +
\tau\hat{\bbox{b}}
\label{fs2}
\end{eqnarray}
where
$\kappa $ is the curvature,
$\tau$ is the torsion
of the curve at arclength position $s$ and time $t$.
The direct evaluation gives
\begin{eqnarray}\nonumber
\frac{ds}{dz} = \sqrt{1+\Sigma_z^{2}},\qquad
\tau = 0,\qquad
\kappa = \frac{\Sigma_{zz}}
{1+\Sigma_z^2}
\end{eqnarray}
Axial symmetry of the problem reduces
evolution to $(\hat{\bbox{n}}, \hat{\bbox{t}})$-plane 
and $(z \leftrightarrow s)$ and
$(\Sigma(z,t) \leftrightarrow \kappa(s,t))$
equivalence.
This describes evolution of an axisymmetric
surface pure geometrically in terms of the curvature and
the arclength.
Integro-differential form of Eqs.~(\ref{hyd-cyl}-\ref{contour})
leads to strong nonlocality as a consequence of long-range
part of the interaction.

\section{Summary}

We consider an uniformly charged incompressible nuclear
fluid bounded by a closed surface.
It is shown that an evolution of an axisymmetric surface
$\Gamma(\bbox{r},t)\equiv \sigma - \Sigma(z,t) = 0,\quad
\bbox{r}=(\sigma,\phi,z)$ can be approximately reduced to
a motion of a curve in the $(\sigma,z)$-plane.
A nonlinear integro-differential equation for the contour
$\Sigma(z,t)$ is derived. It is shown
the direct correspondence between
$\Sigma(z,t)$ and a local curvature, that gives possibility to
use methods of differential geometry to
analyze an evolution of an axisymmetric nuclear surface.

The goal of this short report is to present the principal line
only. The extended paper is in progress.

Work supported in part by Russian Foundation for Basic Research and
Deutsche Forschungsgemeinschaft.

\end{document}